\documentstyle[epsf]{l-aa}
\begin{document}
\thesaurus{6(8.15.1; 8.22.1; 8.22.3; 11.09.1 IC 1613; 11.12.1; 11.19.5)}
\title{Variable stars in nearby galaxies.\thanks{Based on 
observations collected at ESO-La Silla}}
\subtitle{II. Population I and II Cepheids in Field A of IC 1613}
\author{E. Antonello, L. Mantegazza, D. Fugazza, M. Bossi}
\offprints{E. Antonello} 
\institute{Osservatorio Astronomico di Brera, Via E.~Bianchi 46,
       I--23807 Merate, Italy \\
(elio,luciano,fugazza,bossi@merate.mi.astro.it)}
\date{ Received date; accepted date }
\maketitle
\markboth{E. Antonello et al.: Cepheids in IC1613}
{E. Antonello et al.: Cepheids in IC1613}

\begin{abstract}
The light curves of Cepheids and other variable stars in Field A of 
IC 1613, obtained with a CCD and no filter ($Wh$ photometry), have been
analyzed. It is possible to separate first overtone from 
fundamental mode population I Cepheids taking into account the pulsation 
amplitude, the shape of the light curve and the period. The expected 
separation is verified in the period--luminosity $PL$ diagram.
Light curve Fourier parameters have been compared with those of 
Magellanic Clouds and galactic Cepheids, in order to point out the effects of 
the very low metallicity of IC 1613 on the light curve shape. 

Population II Cepheids of IC 1613 can be discriminated from those of
population I in the $PL$ diagram, and, taking into account their color,
from other red or blue variables.  Their $PL$ relation is consistent 
with that observed in globular clusters, nearby dwarf spheroidal galaxies and 
LMC. We have shown it is possible to apply the single-phase method for 
deriving standard photometry $PL$ relations for population I and II Cepheids;
therefore with just one accurate $BVRI$ observation it is possible
to use the population I Cepheids for distance determinations.

Some unusual stars have been identified on the basis of periods, light curve
shapes and colors; they appear to be pulsating stars laying on the extension
of $PL$ relation of known anomalous Cepheids. A firmer classification of
these and other faint stars requires further deeper multicolor observations.

\end{abstract}


\section{Introduction}
Cepheids are important stars not only as primary distance indicators but also
as an essential tool for testing the theories on the internal constitution of 
stars and stellar evolution. The fact that resonances among the pulsation 
modes give rise to observable effects, i.e. structures in the Fourier 
decomposition coefficients, can be exploited to put constraints on the 
pulsational models and on the mass-luminosity relations. In the fundamental 
mode Cepheids a resonance occurs between the fundamental and the second 
overtone mode ($P_0/P_2=2$) in the vicinity of a period $P_0=10$ d and it 
is at the origin of the well known Hertzsprung progression of the bump 
Cepheids (e.g. Simon and Lee \cite{sl}). In the first overtone mode Cepheids, 
a resonance occurs between the first and the fourth pulsation modes 
($P_1/P_4=2$; e.g. Antonello \& Poretti \cite{ap}; Antonello et al. 
\cite{apr}). When these resonances observed in Cepheids of our Galaxy 
and Magellanic Clouds are used to constrain purely radiative models, one 
obtains stellar masses that are too small to be in agreement with stellar 
evolution calculations (e.g. Buchler 1998). It is therefore important to 
observe Cepheids in galaxies with very different metallicities.
The MACHO, EROS and OGLE projects dedicated to the detection of microlensing 
events in the direction of Magellanic Clouds produced enormous amount of data 
on variable stars in these galaxies (e.g. Welch et al. \cite{we}; Beaulieu \& 
Sasselov \cite{bs}; Udalski et al. \cite{ogle1}). 
More recently, the project DIRECT was dedicated to the 
massive CCD photometry of M31 (and M33) with the purpose of detecting Cepheids 
and eclipsing binaries for direct distance determination of these galaxies 
(e.g. Kaluzny et al. \cite{kal}). In the previous paper of this series 
dedicated to the study of variable stars, and particularly Cepheids, in 
nearby galaxies (Antonello et al. \cite{a1}, hereinafter Paper I), we have 
illustrated the methods and strategy adopted for the detection of these stars 
in Field A of IC 1613, and we have shown the usefulness of the observations 
in white-light or $Wh$-band for obtaining the best light curves with 
relatively small telescopes.

The Cepheids of the irregular galaxy IC 1613 [$\alpha=1^h 04^m 50^s$ (2000), 
$\delta= +2\degr 08'$ (2000), l=130\degr, b=--61\degr], were studied by Baade,
but his extensive results were never published. The data, reduced to a new 
photometric scale, were published by Sandage (\cite{san}), who discussed the 
apparently anomalous slope of the $PL$ relation. Freedman (\cite{fre1}), 
Sandage (\cite{sa2}) and Carlson \& Sandage (\cite{cs}) considered again this 
case and the conclusion was that there are no differences in the slope of the 
$PL$ relation of Cepheids in IC 1613 with respect to that of other galaxies.
From CCD $BVRI$ observations (Freedman, \cite{fre1}), Madore \&
Freedman (\cite{mf}) derived a total mean reddening of $E(B-V)$=0.02 mag, and 
a true distance modulus of 24.42${\pm}$0.13 mag, corresponding to a distance 
of 765 kpc; they suggest that IC 1613 is the best place for work on intrinsic 
calibration problems of the Cepheid distance scale, because the foreground 
reddening to this galaxy is very low and probably quite uniform, the 
extinction internal to IC 1613 appears to be quite small and the crowding of 
stellar images are relatively low. IC 1613 has probably a very low average
metallicity, $[Fe/H] \sim -1.3$ (Freedman \cite{fre2}).

In the present paper we discuss the properties of population I
Cepheids observed in Field A of IC 1613, and compare them with those in 
Galaxy and Magellanic Clouds, in a similar way to what done for these
galaxies by MACHO, EROS and OGLE groups, and discuss the population II 
variables. Moreover, we show the utility of $Wh$ light curves for 
deriving a standard $PL$ relation.

\begin{table}
\caption[]{Population I Cepheids}
\begin{flushleft}
\begin{tabular}{lllllll}
\hline\noalign{\smallskip}
 Star  &  $P$ [d] & $Wh_0$ & $V$ & $R$  & $\Delta{Wh}$ & mode\\
\noalign{\smallskip}
\hline\noalign{\smallskip}
 V2396 & 145.6  & 17.43 & 17.99 & 17.45  &   ...  &   F   \\
 V1039 & 16.43  & 19.82 & 20.02 & 19.53  &   .66  &   F  \\ 
 V2414 &  7.573 & 20.83 & 21.11 & 20.57  &   .22  &   F\\
 V1337 &  6.743 & 20.45 & 20.27 & 19.96  &   .76  &   F\\
 V0107 &  6.714 & 20.25 & 20.56 & 19.96  &   .25  &   F\\
 V1734 &  5.737 & 20.68 & 20.64 & 20.21  &   .83  &   F\\
 V2221 &  5.721 & 21.12 & 20.97 & 20.44  &   .53  &   F\\
 V0819 &  5.578 & 21.06 & 21.62 & 21.05  &   .80  &    F\\
 V1592 &  4.360 & 21.76 & 21.69 & 21.33  &  1.30  &   F\\
 V1897 &  4.065 & 20.92 & 20.60 & 20.42  &   .76  &   F\\
 V2256 &  3.663 & 21.29 & 21.84 & 21.42  &  1.00  &    F\\
 V1798 &  3.272 & 21.68 & 22.50 & ...    &   .60  &   F\\
 V1014 &  2.950 & 21.97 & 21.50 & 21.27  &  1.20  &   F\\
 V0555 &  2.868 & 21.49 & 21.70 & 21.44  &   .96  &   F\\
 V0551 &  2.690 & 22.28 & 22.75 & 22.31  &   .60  &    ?\\
 V2150 &  2.567 & 21.47 & 21.17 & 20.83  &   .79  &   F\\
 V0655 &  2.538 & 22.41 & ...   &   ...  &   .69  &   ?\\
 V0414 &  2.459 & 22.27 & ...   &   ...  &   .94  &   F\\
 V1756 &  2.445 & 22.27 & ... &   ...    &  1.03  &   F\\
 V2766 &  2.238 & 21.84 & ... &   ...    &   .90  &   F\\
 V2309 &  2.180 & 21.95 & ... &  21.67   &   .87  &   F\\
 V0279 &  2.098 & 22.46 & 22.32 & 22.13  &   .52  &   F\\
 V2100 &  1.879 & 22.69 & ... &   ...    &   .75  &   ?\\
 V2020 &  1.870 & 22.33 & 22.71 &  ...   &   .33  &   1-O\\
 V0533 &  1.845 & 22.45 & 22.81 & 22.27  &  1.40  &   F\\
 V0377 &  1.643 & 22.23 & 22.41 &  ...   &   .36  &   ...\\
 V1178 &  1.628 & 22.36 & ... &   ...    &  1.12  &   F\\
 V0368 &  1.479 & 22.05 & 22.02 & 21.86  &   .47  &   1-O\\
 V2262 &  1.443 & 21.71 & ... & 21.51    &   .55  &   ...\\
 V0078 &  1.435 & 22.15 & ... &  ...     &   .90  &   F\\
 V2172 &  1.431 & 22.12 & 22.30 & ...    &   .45  &   1-O\\
 V0524 &  1.424 & 22.49 & ... &  ...     &   .40  &   ...\\
 V0236 &  1.390 & 22.26 & ... & 21.95    &   .80  &   ...\\
 V1100 &  1.287 & 22.16 & ... & 21.91    &   .56  &   1-O\\
 V0128 &  1.251 & 22.34 & ... &  ...     &   .59  &   1-O\\
 V1479 &  1.123 & 22.15 & 22.71 & 22.39  &  .32  &   1-O\\
 V1241 &  1.049 & 21.95 & 22.79 & 22.31  &  .53  &   1-O\\
 V1371 &   .886 & 22.38 & ... &  ...     &   .36  &   1-O\\
 V1296 &   .859 & 22.53 & ... &  ...     &   .40   &  ...\\
 V0178 &   .817 & 22.40 & ... & 21.87    &   .54  &   ...\\
 V1767 &   .797 & 22.31 & ... &  ...     &   .35  &   2-O?\\
 V0479 &   .663 & 22.35 & ... &  ...     &   .50  &   ...\\
 V1289 &   .646 & 22.33 & ... &  ...     &   .35  &   ...\\
\noalign{\smallskip}
\hline
\end{tabular}
\end{flushleft}
\end{table}

\section{$Wh$--photometry}
As described in Paper I, the observations were performed at ESO LaSilla
with the 0.9 m telescope and without filter, in order to collect the largest 
possible number of photons. The effective wavelength of $Wh$--band for a 
back--illuminated CCD detector is intermediate between that of Johnson $V$ 
and $R$ bands for F-G spectral types, therefore we expect that the photometric 
characteristics of pulsations such as amplitude and light curve shapes 
(or Fourier parameters) should be correspondingly intermediate between 
$V$ and $R$. In general, the relation between $V$ and $R$ amplitudes
of Cepheids is ${\Delta}R \sim 0.7 {\Delta}V$, the amplitude ratios 
$R_{i1}$ for bands $R$ and $V$ are similar, while the phase differences 
$\phi_{i1}$ are not the same, $\phi_{i1}(R) \sim \phi_{i1}(V)+a$, where 
$a \sim 0.2 - 0.3$ rad (Simon \& Moffett \cite{sm}). Taking into account the 
approximate relation $V-Wh \sim 0.6(V-R)$ derived in Paper I and the relation
between $\Delta{R}$ and $\Delta{V}$, we obtain $\Delta{Wh} \sim 0.8\Delta{V}$.
If we consider the uncertainties and the scatter of parameters from star
to star, we can compare qualitatively the results of $Wh$ photometry,
as regards light curve characteristics, with those of $V$ band. 
Similarly, we will compare our results with those of EROS survey 
(Beaulieu \& Sasselov \cite{bs}), which were obtained with a nonstandard 
$V$-filter, whose effective wavelength was bluer than Johnson $V$-filter.

\section{Population I Cepheids}
\subsection{Characteristics}

The detected population I Cepheids are listed in Table 1, where the 
identification number, period, mean $Wh_0$, single-phase $V$ and $R$ values
(see Paper I), amplitude $\Delta{Wh}$ and pulsation mode have been reported.
The $Wh$ light curves were Fourier decomposed with the formula
\begin{equation}
Wh=Wh_0 + \sum A_i ~ \cos [2{\pi}if(t-T_0) + \phi_i],
\end{equation}
where $f=1/P$; then the Fourier parameters, that is phase differences 
$\phi_{i1}=\phi_i-i\phi_1$ and amplitude ratios $R_{i1}=A_i/A_1$, were 
derived. The decomposition gave reliable results for many Cepheids; the best 
example is shown in Fig. 1, where the light curve of V1337 is reported along 
with the 5th order fit, with a standard deviation of 0.026 mag.
%figure1
\begin{figure}
\epsfxsize=5.6truecm
\epsffile[10 300 380 700]{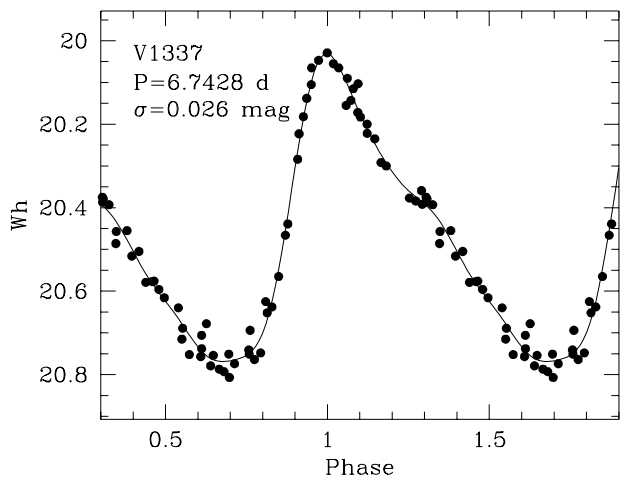}
\caption[ ]{$Wh$ light curve of the Cepheid V1337 with a 5th order Fourier fit 
}  
\end{figure}
%figure2
\begin{figure}
\epsfysize=16truecm
\epsffile[30 170 500 700]{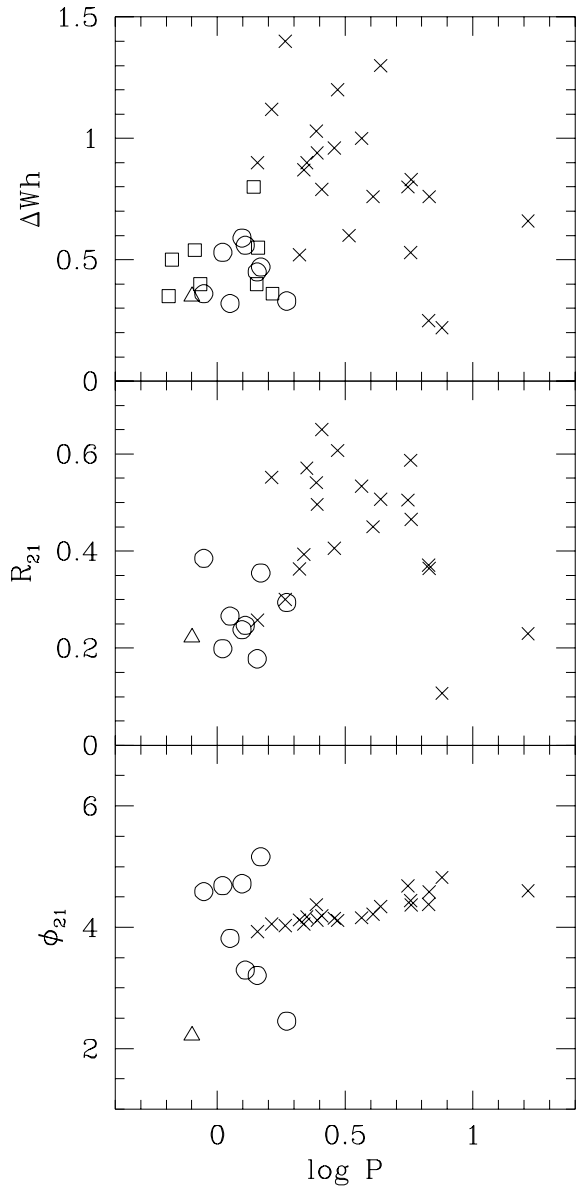}
\caption[ ]{Amplitude $\Delta{Wh}$, $R_{21}$ and $\phi_{21}$ against period 
for population I Cepheids in Field A of IC 1613. {\em Crosses:} fundamental 
mode; {\em open circles:} first overtone mode; {\em open squares:} uncertain 
mode, but most of these stars should be first overtone mode pulsators; 
{\em open triangle:} second overtone mode candidate
}  
\end{figure}
%figure3
\begin{figure}
\epsfxsize=16truecm
\epsffile[40 500 500 700]{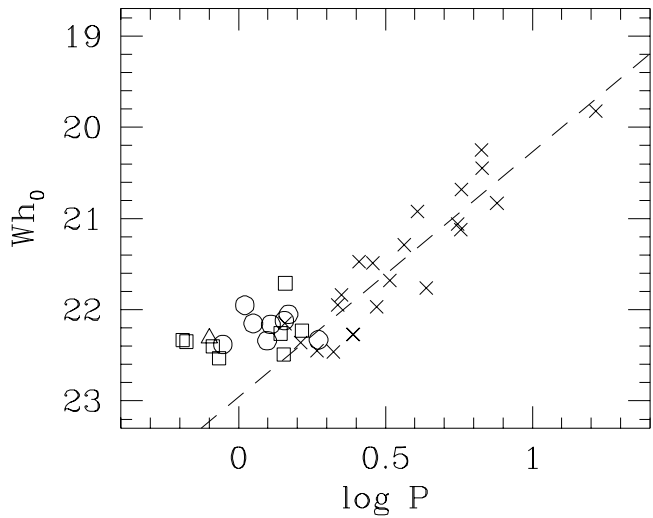}
\caption[ ]{Period--luminosity relation for population I Cepheids in Field A 
of IC 1613. Symbols as in Fig. 2. The dashed line is the statistical relation 
obtained for fundamental mode pulsators
}
\end{figure}
%figure4
\begin{figure}
\epsfxsize=5.6truecm
\epsffile[10 300 380 700]{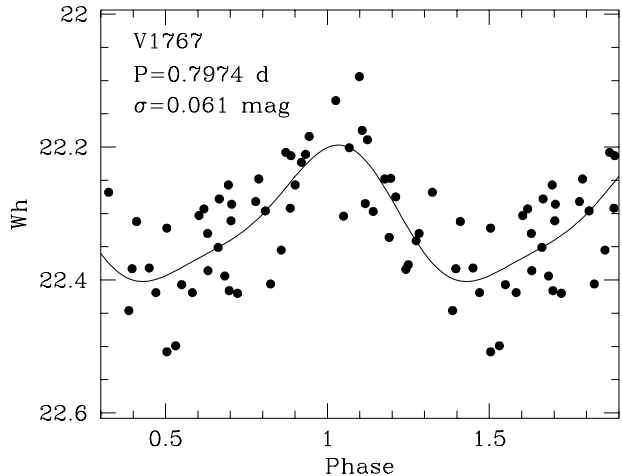}
\caption[ ]{$Wh$ light curve of the second overtone candidate V1767 with a 2nd
order Fourier fit; this figure illustrates also the present limiting 
capabilities of a 0.9 m telescope for discovering small amplitude faint 
Cepheids
}  
\end{figure}
However, for several short period, faint Cepheids the Fourier decomposition 
is too uncertain, and on this basis their mode could not be identified; 
taking into account the amplitude and position in the $PL$ diagram, most of
them should be probable first overtone pulsators. The Fourier parameters 
$R_{21}$, $\phi_{21}$ and the amplitude $\Delta{Wh}$ are plotted against $P$ 
in Fig. 2. The trends of the parameters are qualitatively similar to those 
observed in other galaxies. Unfortunately there are no stars with $P$ 
between 8 and 16 d, and therefore it is not possible to put a constrain with 
just these Cepheids on the location of $P_0/P_2=2$ resonance; this point will 
be discussed also in the next section. As regards first overtone mode Cepheids,
we have to remark the apparent lack of stars with $P > 2$ d (log $P>$ 0.3). 
We have looked carefully at the data base, but we have not found convincing 
candidates with such periods. First overtone Cepheids are characterized by a 
scatter of $\phi_{21}$ values, which could be interpreted as the effect of
the resonance $P_1/P_4=2$. The $\phi_{21}$ values are uncertain,
with a formal error of about 0.5 rad, and therefore we cannot put much weight 
on this result.

First overtone mode pulsators are clearly discriminated also in the $PL$ 
diagram shown in Fig. 3; indeed they are located roughly about half a 
magnitude above the $PL$ relation of fundamental mode Cepheids,
\begin{equation}
Wh_0=-2.69 \log P + 22.96.
\end{equation}

There is at least one second overtone candidate, V1767; the light
curve, shown in Fig. 4, is different from that of other Cepheids,
and we suspect it is the signature of the expected resonance 
$P_2/P_6=2$ for $P \la 1$d. This result would be consistent with theoretical
expectations (Antonello \& Kanbur \cite{ak}) and with the observational results
for LMC double-mode Cepheid pulsating in first and second overtone modes
(Alcock et al. \cite{macho3}). The search for pure second overtone mode 
Cepheids in SMC (e.g. Mantegazza \& Antonello \cite{ma}) has yielded 
only recently the first positive results in the context of OGLE experiment 
(Udalski et al. \cite{ogle1}). According to the results of OGLE group, 
the SMC second overtone mode Cepheids have very small amplitudes ($<0.2$ mag)
and the light curves are almost sinusoidal; taking into account the scatter,
V1767 has an amplitude of about  0.2 -- 0.3 mag,  but the shape is a bit
different from that of a sinusoid.

%figure5
\begin{figure}
\epsfysize=17truecm
\epsffile[40 180 380 700]{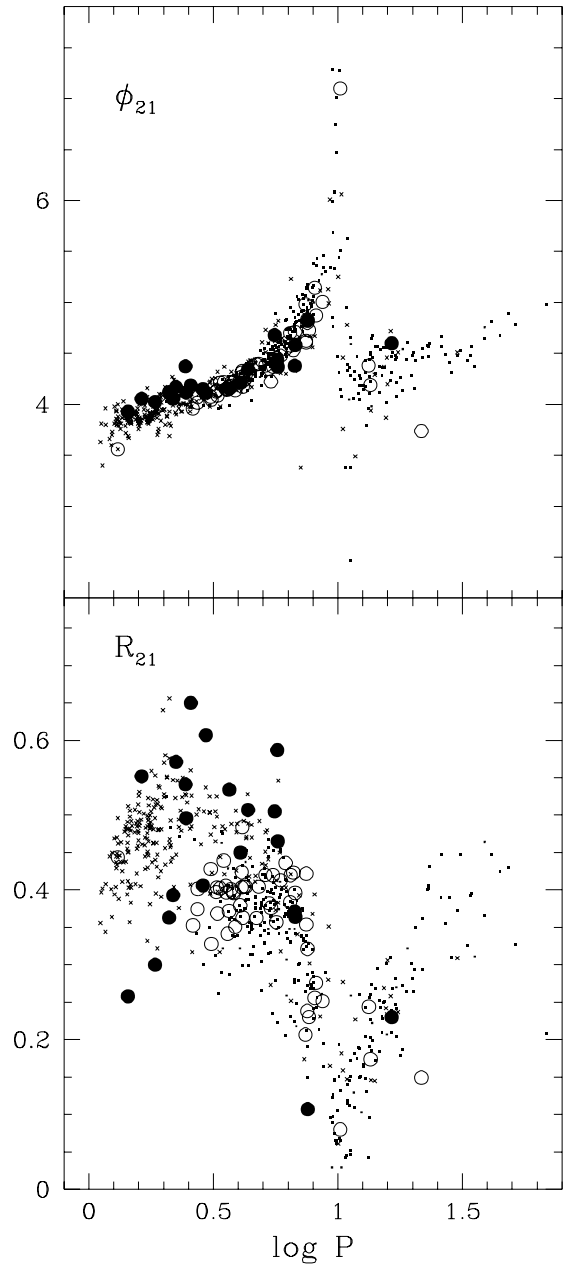}
\caption[ ]{Phase difference $\phi_{21}$ and amplitude ratio $R_{21}$ 
against period for fundamental mode Cepheids in Galaxy ({\em dots}), 
LMC ({\em open circles}), SMC ({\em crosses}) and IC 1613 
({\em filled circles})
}
\end{figure}
%figure6
\begin{figure}
\epsfxsize=12truecm
\epsffile[50 410 380 700]{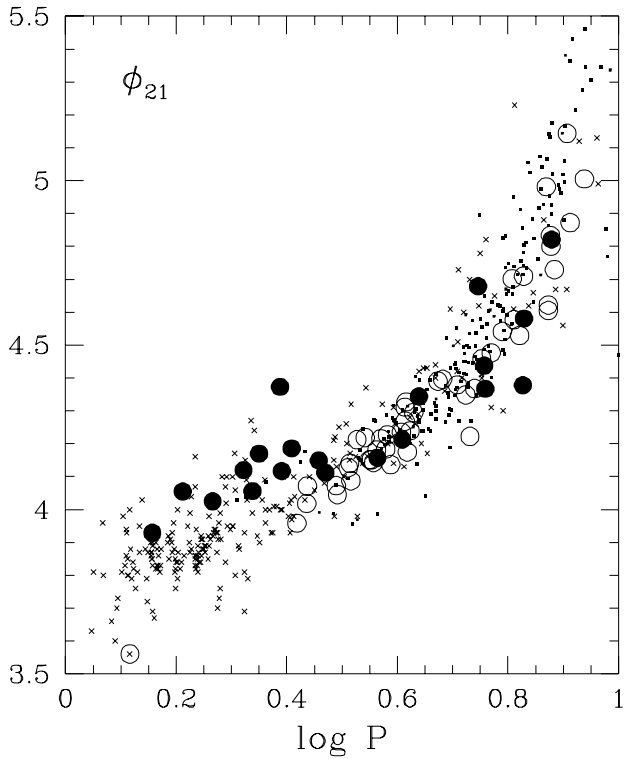}
\caption[ ]{Detail of the $\phi_{21}$--$P$ diagram shown in Fig. 5
}
\end{figure}

Two warnings are in order: the first concerns the colors of the stars
and the second the aliasing problem. Since $Wh$ photometry is deeper than $V$
and $R$ ones made with the same telescope, presently we have no information 
on the colors of the faintest Cepheids. In particular, as noted by Udalski 
et al. (\cite{ogle1}), small amplitude, almost sinusoidal light curves
can be produced by several objects, such as ellipsoidal variables (which
are usually hot stars) or spotted stars (which are usually cool stars).
As remarked in Paper I, there are some difficulties in identifying the 
correct frequency for shorter period variables, because the peaks in the power 
spectrum at $f$ and at the alias $f-1$ are similar; for example, 
a solution with $P=3.9$d (but with a lower power spectrum peak) for V1767
is also reasonable, but it is too long a period for a population I Cepheid 
with the same luminosity, and the light curve keeps a non-sinusoidal shape.
We conclude that shorter period Cepheids must be considered good candidates 
at least until a better information on their color is available.

The amplitudes as a function of P shown in Fig. 2 (upper panel) indicate a 
distribution similar to that observed in Galaxy. In particular, the first 
overtone mode pulsators have generally small amplitudes.

\subsection{Comparison with other galaxies}
Interesting results are obtained when we compare directly the Fourier
parameters of Cepheids in IC 1613, Galaxy (data from Mantegazza \& Poretti
\cite{mp}, Antonello \& Morelli \cite{am}, and references therein) 
and Magellanic Clouds (EROS data; Beaulieu \& Sasselov 
\cite{bs}). $\phi_{21}$ and $R_{21}$ are plotted against the period in Fig. 5. 
It is quite evident that $\phi_{21}$ values of fundamental mode Cepheids 
are very similar in all the galaxies; this result, coupled with the evidence 
that $R_{21}$ has a minimum for the same period range, confirms clearly that 
the resonance effects of $P_0/P_2=2$ do not depend much on the metallicity. 
Even if we noted previously that IC 1613 data alone cannot put a constraint, 
they tend to follow the same trend as that of the other galaxies. Fig. 6 shows
a detail of the $\phi_{21}$ -- $P$ diagram. The fact that the phase difference
appears to be not very sensitive to metallicity, and in a certain sense nor to
the passband, is striking. The Cepheids of Galaxy have $[Fe/H]\sim 0.0$ and 
the Fourier parameters were obtained for $V$ light curves; for LMC and SMC 
($[Fe/H]\sim -0.2$ and $[Fe/H]\sim -0.5$, respectively; Feast 1991) the 
passband was bluer than $V$, while for IC 1613 ($[Fe/H]\sim -1.3$) the 
passband was redder. The uniformity of $\phi_{21}$ values is not likely
to be due to curious combined effects. The clearest indication 
for a small difference is the trend of IC 1613 Cepheids with shorter $P$: the
$\phi_{21}$ values tend to be slightly larger than those for the other 
galaxies. It may be possible that the $\phi_{21} - P$ relation in the 
range $1 \la P \la 8$ d ($0 \la \log P \la 0.9$) is a function of the 
metallicity.

One should note the systematic differences of $R_{21}$ envelopes for shorter 
$P$ among the various galaxies; as remarked in Sect. 2 this does not depend
on the different photometric band, but it should reflect the sensitivity
of this parameter to the metallicity. It may be possible that IC 1613 Cepheids
have the largest $R_{21}$ values because they are the metal poorest. 
Since the $\phi_{21}$ values do not change too much from galaxy to galaxy, 
increasing $R_{21}$ values for a given period range are related to light 
curves with steeper rising branches and possibly also stronger humps 
and bumps. For the same reason the amplitudes of IC 1613 Cepheids should be
quite larger than for example those of Galaxy.
%figure 7
\begin{figure}
\epsfysize=17truecm
\epsffile[40 180 380 700]{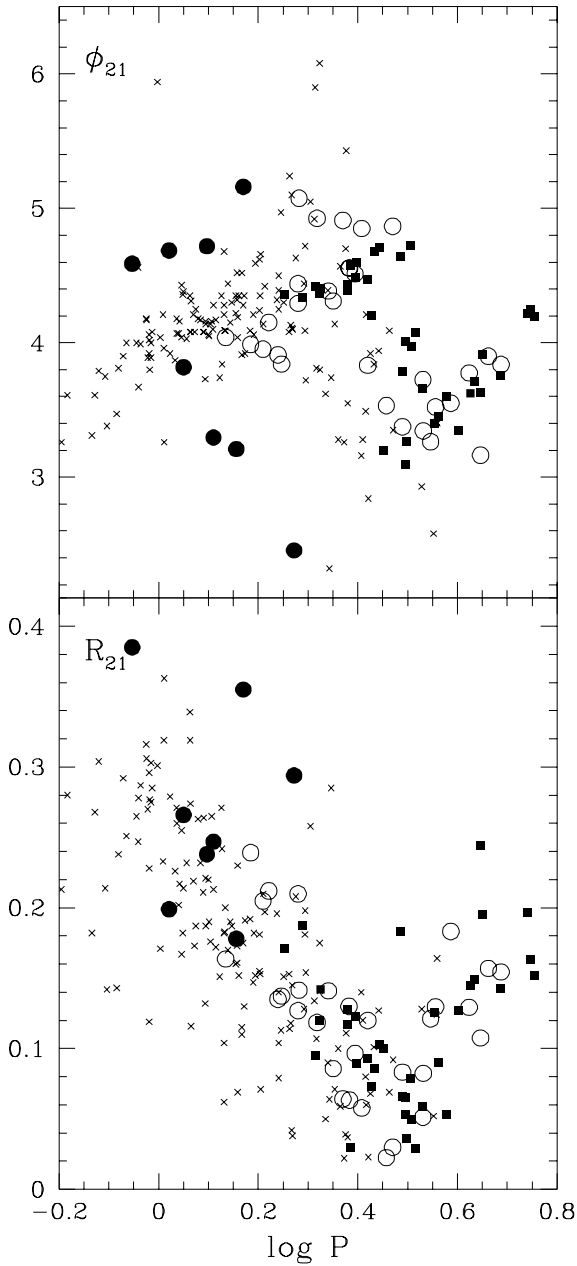}
\caption[ ]{Phase difference $\phi_{21}$ and amplitude ratio $R_{21}$ 
against period for first overtone mode Cepheids in Galaxy ({\em filled 
squares}), LMC ({\em open circles}), SMC ({\em crosses}) and IC 1613 
({\em filled circles})
}
\end{figure}
%figure8
\begin{figure}
\epsfysize=11truecm
\epsffile[30 340 380 690]{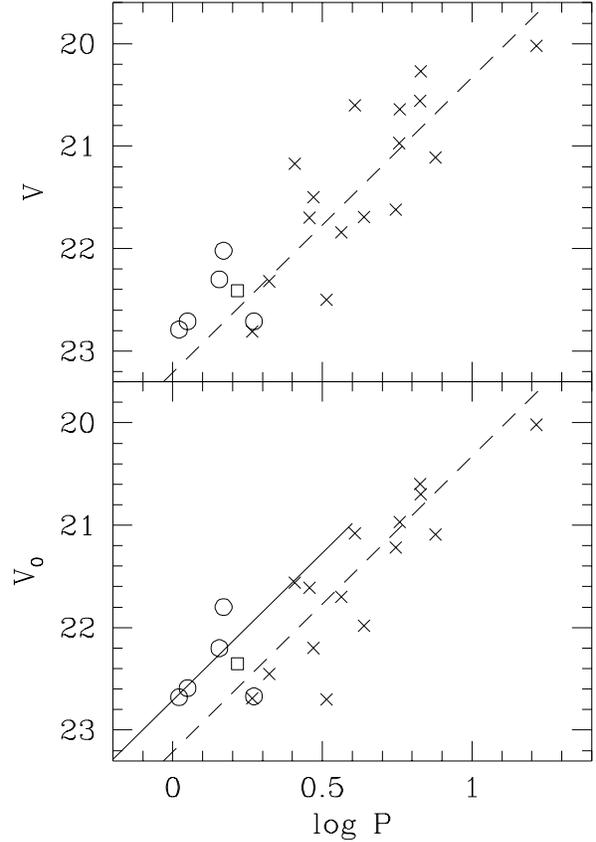}
\caption[ ]{$PL$ relations with $V$ data. Upper panel: random-phase 
single-value relation. Lower panel: converted to mean $V$ light relation 
(see text). {\em Crosses:} fundamental mode
Cepheids; {\em open circles:} first overtone mode Cepheids; {\em open
square:} Cepheids with uncertain mode. {\em Dashed line:} $PL$ relation for 
$V$ band of 25 LMC stars (Madore \& Freedman \cite{mf}) plotted assuming
apparent values for distance modulus $\mu=24.45$ for IC 1613 
and relative distance $\delta\mu=5.71$ between IC 1613 and LMC (Freedman
\cite{fre1}). {\em Continuous line:} relation with the same slope and 
decreased zero-point by 0.5 mag, indicating approximatively the expected 
relation for first overtone mode Cepheids
}  
\end{figure}

The first overtone mode Cepheid case is shown in Fig. 7. $R_{21}$ values 
appear to follow the same trend in Galaxy and Magellanic Clouds
as in the fundamental mode case, and also IC 1613 data, even if uncertain, are
roughly consistent with this picture. The $P$ corresponding to the minimum of 
$R_{21}$ values for Galaxy and Magellanic Clouds tends to differ slightly from
galaxy to galaxy, that is it decreases with the metallicity. It is 
possible to identify approximately the following period of minimum: 3.2 d 
(Galaxy), 2.9 d (LMC) and 2.5 d (SMC). The $\phi_{21}$ values appear even 
more sensitive to the different metallicity. The so-called Z--shape of 
$\phi_{21}$ values in the Galaxy is progressively distorted for decreasing 
metallicity. The lowest  $\phi_{21}$ values at $\log P\sim 0.5$ tend to 
coincide in Galaxy and Magellanic Clouds, while the highest $\phi_{21}$ 
values are displaced towards the shorter periods for decreasing metallicity. 
On the basis of this feature the EROS group
(e.g. Beaulieu \& Sasselov \cite{bs2}) remarked a change of the center of the 
resonance effect $P_1/P_4=2$ as a function of metallicity. 
The intriguing question now is 
whether in IC 1613 the first overtone values indicate the resonance effect at 
much shorter period or it is just a matter of uncertain data. Could the large 
metallicity difference be responsible for this? Better data are needed to 
solve this problem.

%figure9
\begin{figure}
\epsfysize=6truecm
\epsffile[30 500 380 690]{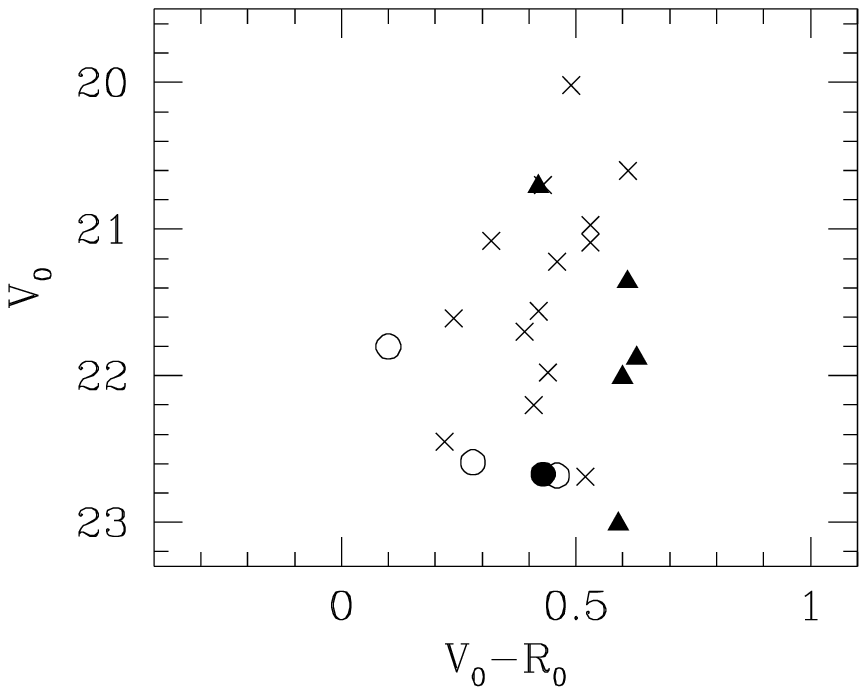}
\caption[ ]{Color--magnitude diagram for population I and II Cepheids.
The $V$ and $R$ values are those converted to mean light. 
{\em Crosses:} fundamental mode pop I Cepheids; {\em open circles:} 
first overtone pop I Cepheids; {\em filled triangles:} 
W Vir stars; {\em filled circle:} possible anomalous Cepheid
}  
\end{figure}
%figure10
\begin{figure}
\epsfysize=11truecm
\epsffile[30 340 380 690]{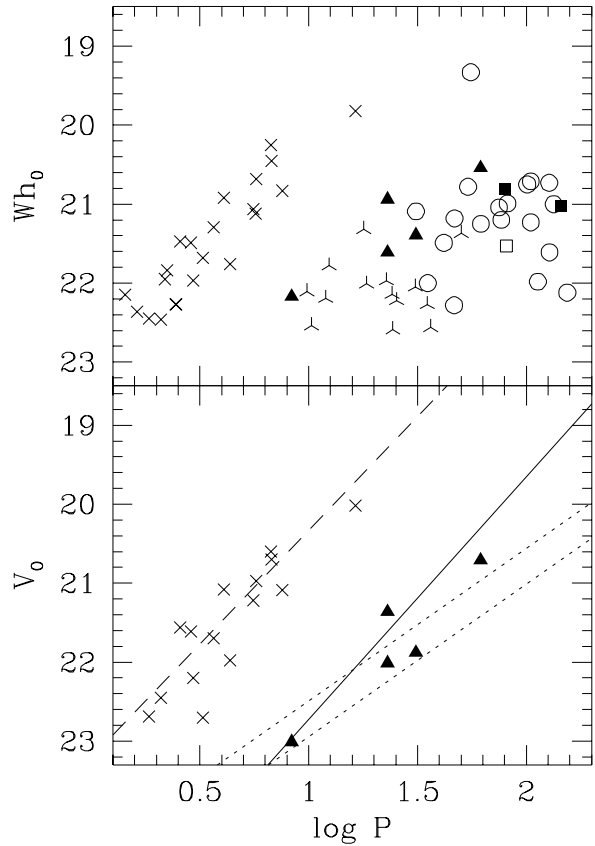}
\caption[ ]{$PL$ relations. Upper panel: $Wh_0$ data. Lower panel: 
converted to mean $V$ light relation. {\em Crosses:} fundamental mode 
Cepheids shown in Fig. 8;  {\em filled triangles:} W Vir stars; 
{\em skeletal triangles:} other W Vir and periodic variables with unknown
color; {\em filled squares:} possible RV Tau stars; {\em open circles:} 
semiregular and long period variables; {\em open squares:} variables 
with $V$--$R \la 0.1$. {\em Dashed line:} $PL$ relation for classical 
Cepheids shown in Fig. 8; {\em dotted lines:} $PL$ relations for 
population II Cepheids pulsating in the fundamental (lower line) and 
first overtone mode (upper line) derived by Nemec et al. (\cite{nem}); 
{\em continuous line:} relation 
obtained by Alcock et al. (\cite{macho1}) for LMC stars
}  
\end{figure}

\section{Single-phase $PL$ relation}
In the present section we discuss the possibility of the application of 
the single-phase method devised by Freedman (\cite{fre1}), that is we will 
use the only available $V$ observation of the (brightest) Cepheids and the 
characteristics of $Wh$ light curves. Cepheid light curve shapes and 
amplitudes depend on the color or passband as mentioned in Sect. 2. We take 
into account the ratio $r$=0.8 between $Wh$ and $V$ amplitudes and for the 
present we will ignore the presumably small phase shift of maximum between 
$Wh$ and $V$ light curves. In principle, simultaneous observations of a test 
Cepheid are needed in order to derive the accurate ratio between $Wh$ and $V$ 
amplitudes and the phase shift; however, considering the present precision of 
data, we think this is not a strict requirement. In particular, the error in 
the $V$ measurement of faintest Cepheids estimated by DAOPHOT is large, 
$\ga 0.18$ mag. A detailed study (also with simulations) of this topic will 
be made when all the observations of the four fields of IC 1613 will be 
reduced. For the present, we show the result of our exercise in Fig. 8; 
the mean $V$ magnitude of Cepheids was estimated with the formula
\begin{equation}
V_0 = V(\phi)-[Wh(\phi) - Wh_0]/r,
\end{equation}
where $V(\phi)$ is the single measurement, $Wh(\phi)$ the value of the 
fitted $Wh$ curve for the same phase and $r$=0.8. Actually, $V_0$ is a 
converted single-phase $V$ value to mean value. The comparison of the two 
panels in Fig. 8 shows the expected improvement of the converted value over 
the random single--phase relation mainly for the brighter Cepheids; the 
expected approximate relation for first overtone mode pulsators is also 
plotted. For deriving $R_0$ from the single-phase $R$ value we have adopted
the same procedure using $r=0.6$; both $V_0$ and $R_0$ values are
listed in Table 3.

We have planned to make observations in $BVRI$ bands of the four observed 
fields of the galaxy with a larger telescope in order to obtain {\em one} 
accurate photometric value in each of these bands for all of the Cepheids; 
this will allow us to derive statistically significant standard $PL$ 
relations for fundamental and first overtone mode pulsators using the 
single--phase method.

\section{W Vir stars and other population II Cepheids}
W Vir stars can be discriminated from population I Cepheids in the $PL$ 
diagram, because the former, for a given mean magnitude, have longer periods.
In our Galaxy, W Vir stars are bluer than classical Cepheids, while in the 
very metal poor galaxy IC 1613 the color difference between the two 
populations presumably is small. Therefore we have identified the 
probable W Vir stars in the sample of the longer period variables with 
Cepheid--like light curve on the basis of their color; they are reported in 
Table 2 along with their mean $Wh$ value and single-phase $V$ and 
$R$ values. Converted to mean light curve values were obtained 
with the same procedure adopted for population I Cepheids, and are listed
in Table 3.
Fig. 9 shows the color-magnitude diagram for Cepheids of both populations;
the $V$--$R$ color range is 0.10 -- 0.63 mag, to be compared with the
range indicated by the 11 population I Cepheids observed by Freedman
(\cite{fre1}), 0.15 -- 0.66 mag.
\begin{table}
\caption[]{W Vir stars}
\begin{flushleft}
\begin{tabular}{llllll}
\hline\noalign{\smallskip}
Star & P [d]  & $Wh_0$  & $V$ &  $R$ & $\Delta{Wh}$ \\
\noalign{\smallskip}
\hline\noalign{\smallskip}
 V1935  &  61.7   & 20.54 & 20.90 & 20.43 & 0.8  \\
 V0971  &  31.15  & 21.39 & 21.94 & 21.30 & 0.2  \\
 V1598  &  23.01  & 20.94 & 21.36 & 20.75 & 0.5  \\ 
 V0130  &  22.99  & 21.61 & 21.93 & 21.35 & 0.3  \\
 V0881  &  8.353  & 22.17 & 22.74 & 22.23 & 0.8  \\
\noalign{\smallskip}
\hline
\end{tabular}
\end{flushleft}
\end{table}

%figure11
\begin{figure}
\epsfysize=6truecm
\epsffile[30 500 380 690]{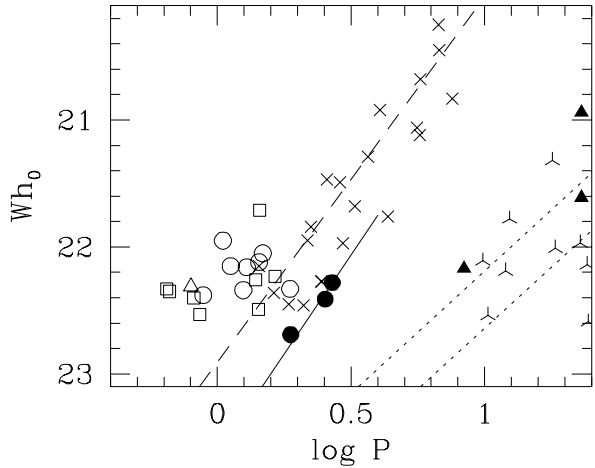}
\caption[ ]{Period--$Wh_0$ diagram for faint Cepheids. 
{\em Crosses:} fundamental mode population I Cepheids; {\em open circles:} 
first overtone population I Cepheids; {\em open squares:} population I 
Cepheids of uncertain mode; {\em open triangle:} possible second overtone 
population I Cepheid; {\em filled triangles:} W Vir stars; {\em skeletal 
triangles:} other periodic variables; {\em filled circles:} possible anomalous 
Cepheids. The lines are the $PL$ relations for the $V$ band corrected
for the color difference $Wh=V-0.3$ (rough mean value for Cepheids);
the continuous line is the extension of the  relation for fundamental mode 
anomalous Cepheids and the dotted lines for first and fundamental mode W Vir
stars derived by Nemec et al. (\cite{nem})
}  
\end{figure}
The W Vir stars are reported in the $PL$ diagram shown in Fig. 10.
In the lower panel the two parallel dotted lines are the $PL$ relations
obtained by Nemec et al. (\cite{nem}) for population II Cepheids in 
Galaxy and nearby spheroidal galaxies, corrected for the distance
modulus 24.45; the lower line is for fundamental mode stars, and the upper 
line for first overtones. The relations take into account the metallicity 
of the galaxies. The continuous line is the relation for population II
LMC stars, $M_V=1.34-3.07\log P$, obtained by Alcock et al. (\cite{macho1}). 
If we consider the various uncertainties in our derivations, there is a 
qualitative agreement. A quantitative analysis requires however a
larger sample of stars of IC 1613. Carlson \& Sandage (\cite{cs})
suggested that the separation between population I and II Cepheid 
relations in IC 1613 is smaller than that in other galaxies; the suggestion 
was based on one Cepheid--like star which was assumed to be a W Vir star. 
Our results show however that in IC 1613 such separation appears to be normal.
\begin{table}
\caption[]{Converted to mean light curve $V$ and $R$ data for population I 
and II Cepheids}
\begin{flushleft}
\begin{tabular}{lllllll}
\hline\noalign{\smallskip}
 Star  &  $V_0$ & $R_0$ &~~~ & Star  &  $V_0$ & $R_0$ \\
\noalign{\smallskip}
\hline\noalign{\smallskip}
 V1039 & 20.02 & 19.53  & &  
   V2020 & 22.67 &  ...     \\
 V2414 & 21.09 & 20.56  & &
   V0533 & 22.69 & 22.17    \\
 V1337 & 20.70 & 20.27  &  &
   V0377 & 22.35 &  ...     \\
 V0107 & 20.60 & 19.99  &  &
   V0368 & 21.80 & 21.70    \\
 V1734 & 20.97 & 20.44  &  &
   V2262 &  ...  & 21.65   \\
 V0819 & 21.22 & 20.76  &  &
   V2172 & 22.20 & ...      \\
 V1592 & 21.98 & 21.54  &  &
   V0236 &  ...  & 22.13    \\
 V1897 & 21.08 & 20.76  &  &
   V1100 &  ...  & 21.76    \\
 V2256 & 21.70 & 21.31  &  &
   V1479 & 22.59 & 22.31    \\
 V1798 & 22.70 &  ...   &  &
   V1241 & 22.68 & 22.22    \\
 V1014 & 22.20 & 21.79  &  &
   V0178 &   ... & 21.79    \\
 V0555 & 21.61 & 21.37  &  &
         &       &          \\
 V0551 & 22.67 & 22.24  &  &
   V1935  &  20.71 & 20.29   \\
 V2150 & 21.56 & 21.14  &  &
   V0971  &  21.88 & 21.25   \\
 V2309 &   ... & 21.73  &  &
   V1598  &  21.36 & 20.75   \\ 
 V0279 & 22.45 & 22.23  &  &
   V0130  &  22.01 & 21.41   \\
       &       &        &  &
   V0881  &  23.01 & 22.42   \\
\noalign{\smallskip}
\hline
\end{tabular}
\end{flushleft}
\end{table}

In Paper I we have mentioned three stars with relatively symmetric light curve,
and with relatively low luminosity and long period. They are too faint for
being first overtone pulsators, and cannot be fundamental mode pulsators
owing to the different light curve shape. Only one of these stars has known
$V$--$R$, which indicates a location in the instability strip (Fig. 9), and 
we suggested tentatively they are anomalous Cepheids. In Fig. 11 it is shown 
the $PL$ diagram for faint Cepheids, which include these stars.
Since the $V$ measurement is lacking for several faint stars, we used
the mean $Wh$ magnitude. The $PL$ relations for the $V$ band adopted for 
classical Cepheids, and for the anomalous Cepheids and W Vir stars 
(Nemec et al. \cite{nem}) were corrected for the mean color of
Cepheids, $Wh=V-0.3$, and the adopted distance modulus was the same
as before (24.45). The result for the anomalous Cepheids is intriguing 
indeed. However, there are some difficulties in accepting this 
interpretation, since there are not known anomalous Cepheids with 
$\log P > 0.2$ in galactic globular clusters and dwarf spheroidal galaxies. 
If existent, such 'long' period stars would have probably similar 
masses to the other pop I Cepheids, and it would be difficult to discriminate
them. Another point is the lack of globular clusters in
IC 1613 (e.g. Hodge \cite{h1}, \cite{h2}). The region of Cepheids with short 
$P$, say $\la 3$ d (log $P \la$ 0.5), appears to be populated by interesting 
and yet unexplained stars: Alves et al. (MACHO group; \cite{macho2}) 
found in LMC a sequence of pulsators which are fainter than that of 
fundamental mode population I Cepheids; our stars could be considered the 
first overtone analogue of such a sequence, assuming that the LMC faint stars 
are fundamental mode pulsators.

\section{Conclusion}
The analysis of the light curve of the Cepheids in Field A of IC 1613 has 
yielded the following results: a) a group of first overtone mode has 
been separated from fundamental mode Cepheids for the first time in a 
galaxy located beyond the Magellanic Clouds; b) a second overtone candidate 
has been identified taking into account its period and unusual light curve 
shape, which could be explained by the expected resonance $P_2/P_6=2$; 
c) conclusions on the effects of a very low metallicity on the light curve 
shape have been drawn from the comparison with Cepheids in other galaxies, 
i.e. Galaxy, LMC and SMC; d) there are no large metallicity effects on the 
resonance $P_0/P_2=2$ for fundamental mode stars, and there are probable 
differences in the resonance $P_1/P_4=2$ for first overtone mode pulsators;
e) the $Wh$ light curves can be confidently used for deriving a standard
$PL$ relation for the $V$ band by applying the single-phase method;
f) the position in the $PL$ diagram of population II Cepheids is consistent 
with that of Cepheids in globular clusters, nearby dwarf spheroidal galaxies 
and LMC; g) some unusual stars have been identified on the basis of the period,
light curve shape and color; they appear to be pulsating stars laying on 
the extension of $PL$ relation of known anomalous Cepheids. 

A firmer classification of the faint variable stars detected in the $Wh$ band
requires complementary deeper multicolor observations; these observations are
needed also for applying the above quoted single-phase method to the faintest
detected Cepheids.

\begin{acknowledgements}
Thanks are due to J.P. Beaulieu (EROS team) for supplying us with
the parameters of SMC and LMC Cepheids.
\end{acknowledgements}

{}
\end{document}